# Observation of the unidirectional magnetoresistance in antiferromagnetic insulator $Fe_2O_3$/Pt bilayers

Yihong Fan[1], Pengxiang Zhang[2], Jiahao Han[2,ς], Yang Lv[1], Luqiao Liu[2], and Jian-Ping Wang[1,*]

[1]*Department of Electrical and Computer Engineering, University of Minnesota, 200 Union St. SE, Minneapolis, Minnesota 55455, USA*

[2]*Department of Electrical Engineering and Computer Science, Massachusetts Institute of Technology, 77 Massachusetts Ave. Room 38-401 Cambridge, MA 02139*

[ς]*Current address: Research Institute of Electrical Communication, Tohoku University, Sendai 980-8577, Japan*

*Corresponding author: jpwang@umn.edu

**Abstract**:

Unidirectional magnetoresistance (UMR) has been observed in a variety of stacks with ferromagnetic/spin Hall material bilayer structures. In this work, we reported UMR in antiferromagnetic insulator $Fe_2O_3$/Pt structure. The UMR has a negative value, which is related to interfacial Rashba coupling and band splitting. Thickness-dependent measurement reveals a potential competition between UMR and the unidirectional spin Hall magnetoresistance (USMR). This work revealed the existence of UMR in antiferromagnetic insulators/heavy metal bilayers and broadens the way for the application of antiferromagnet-based spintronic devices.

The unidirectional spin Hall magnetoresistance (USMR) has attracted a great deal of attention due to its fascinating physics which reveals spin states and electron and magnon transportation properties, as well as application potential in two-terminal reading [1-9]. Compared to the spin Hall magnetoresistance (SMR),

USMR has an angular dependence of 360° rather than 180°, which can reveal the spin-up and spin-down states of the magnetic layer. USMR is first explained with the interfacial spin accumulation and effective current-in-plane giant magnetoresistance [1-3], which is related to the spin Hall effect. Recent research also observed unidirectional magnetoresistive behaviors with different origins, where some of them are spin Hall related and some of them are not. In this case, a more general term of unidirectional magnetoresistance (UMR) is used. Multiple models have been proposed to explain the existence of the UMR and USMR, including spin momentum locking [10,11], magnon contributions [2,12-15], spin-flip contributions [2,13], and spin torque [9].

The spin-orbit torque (SOT) switching of antiferromagnetic material has drawn attention due to its potential in storage applications [16-22]. However, lacking net moment also increases the difficulty to read the switching of the antiferromagnetic order. Conventional reading techniques require 8 or 4 terminals, which significantly increases the complexity and may be debatable because of the thermal-induced magnetoelastic effects [23,24]. Two terminal reading technique is highly demanded in antiferromagnetic spintronics. As the most common two-terminal reading technique, UMR or USMR can be a prospective candidate. However, UMR and USMR have been observed mostly in metallic ferromagnetic systems [1-11] as well as ferromagnetic insulators [15], which reveals their relationship with spin states. Recently, the observation of UMR in antiferromagnetic FeRh/Pt bilayer has paved the way to two-terminal reading in antiferromagnetic systems [25]. Current evidence points out that the potential origin of antiferromagnetic UMR is from interfacial Rashba coupling with the enhancement induced by spin canting [23], which should be confirmed with the existence of UMR or USMR in antiferromagnetic insulator/heavy metal bilayers.

α-$Fe_2O_3$ is a well-studied antiferromagnetic insulator with room temperature antiferromagnetic phase and strong antiferromagnetic exchange interaction [26,27]. Due to the small net magnetization induced by the Dzyaloshinskii-Moriya (DM) interaction which induces a weak anisotropy [27], a low spin-flop field ($<\sim 1T$) is observed in α-$Fe_2O_3$, which provides a platform for current-induced antiferromagnetic switching studies [20-23]. In this work, we observed UMR in antiferromagnetic $Fe_2O_3$/Pt bilayers. The existence of UMR in antiferromagnetic insulators provides a way for two-terminal antiferromagnetic reading and indicates an interfacial origin of the UMR signal.

The α-Fe$_2$O$_3$ films (60 nm) were grown on α-Al$_2$O$_3$ (0001) substrate by magnetron sputtering and post-annealing, which is an easy-plane antiferromagnet according to our previous report [23]. The growing methods and film qualities are identical to our previous works [22,23,28]. As shown in Figure 1(a), the clear X-ray diffraction (XRD) peak of α-Fe$_2$O$_3$ is observed for the α-Fe$_2$O$_3$ (60 nm) sample, suggesting epitaxial growth. The α-Fe$_2$O$_3$ (60 nm) sample was then measured with a SQUID magnetometer at 300 K as shown in Figure 1(b) (easy axis) and 1(c) (hard axis), where a small saturation magnetization of ~2 emu/cm$^3$ is observed, as shown in Figure 1(c). Due to the substrate miscut [28], there is a growth-induced weak uniaxial anisotropy in the easy plane (C-plane), making the three sides of the sample triangle magnetically inequivalent. There is one "easy" side and two other equivalent "hard" sides when we look at the M-H loops measured along the sides. The "hard" side M-H loop has a saturation field of 6000 Oe. The Fe$_2$O$_3$ observes a saturation field of ~6000 Oe at the hard axis [$\bar{1}\bar{1}20$] and [$2\bar{1}\bar{1}0$] directions. 5 nm Pt was grown on the α-Fe$_2$O$_3$ film as a spin Hall channel with magnetron sputtering. To confirm whether the sample is patterned along the easy axis or hard axis, Hall voltage is measured for the sample with magnetic field sweeping along the Hall direction, which is parallel to one of the three sides. The result is shown in figure 2(a), which shows a saturation field of ~6000 Oe. The side that is perpendicular to the current direction is one of the two hard axises of the Fe$_2$O$_3$ layer. To be consistent with previous research, we used [$2\bar{1}\bar{1}0$] as the hard axis direction. The α-Fe$_2$O$_3$/Pt stack was then patterned into a Hall bar (Hall channel size ~35 μm×10 μm), with Hall channel direction parallel to the hard axis [$2\bar{1}\bar{1}0$] direction and current flow towards [$01\bar{1}0$] direction, as shown in the left part of Figure 2(b). The etching removed all the α-Fe$_2$O$_3$ film to ensure the absence of nonlocal magnon transport which may influence the measurement [28]. However, this long etching may also result in a thinner channel length due to sidewall etching, thus the geometry factors were recharacterized. The patterned device was then measured in a physical property measurement system (PPMS) at 300 K with two SR830 lock-in amplifiers and a Keithley 2182A current source, as shown in the right part of Figure 2(b). To avoid the influence of the substrate miscut induced easy and hard axes, a field that is larger than the hard axis saturation field (~6000 Oe) is always used during the measurement.

The second harmonic method is used to characterize the UMR and USMR signals [1-3,5-9]. A 5 mA AC current with a frequency 133.3 Hz is applied to the device. The first and second harmonic signals are measured with two lock-in amplifiers. The first harmonic signal contains information of SMR, which is shown in figure 2(c), and the second harmonic signal contains information of UMR and SOT contributions. The measured second harmonic signals under different magnetic fields are shown in Figure 2(d), where the peak-to-peak signal at 90° and 270° increases with the field. The total signal can be separated into three parts, the contribution of potential UMR, the thermal contribution, and field-like torque contributions. As shown in Figure 2(e) (2(f)), the longitudinal (transverse) angular dependent second harmonic signal is fitted with the following equation [15,29]:

$$V_{xx} = \frac{V_{x,\text{SMR}} H_{\text{FL}}}{2 H_{\text{ext}}} \sin(2\theta) \cos(\theta) + (V_{x,\text{thermal}} + V_{\text{UMR}}) \sin(\theta) \quad (1),$$

$$V_{xy} = \frac{V_{y,\text{SMR}} H_{\text{FL}}}{2 H_{\text{ext}}} \cos(2\theta) \cos(\theta) + V_{y,\text{thermal}} \cos(\theta) \quad (2),$$

where $V_{xx}$ ($V_{xy}$) is the total measured signal in the longitudinal (transverse) direction, $V_{x,\text{SMR}}$ ($V_{y,\text{SMR}}$) is the SMR voltage which is related to the contribution of field-like torque, $H_{\text{FL}}$ is the field-like torque effective field, $H_{\text{ext}}$ is the external field, $\theta$ is the angle between current and field, $V_{x,\text{thermal}}$ ($V_{y,\text{thermal}}$) is the thermal contribution at longitudinal (transverse) direction which includes anomalous Nernst effect (ANE) and spin Seebeck effect (SSE)[30], and $V_{\text{UMR}}$ is the contribution of UMR, respectively. The reference sample (5nm Pt on $Al_2O_3$ substrate) is also measured with the same field and current values, and the result is shown in Figure 2(e) (grey dots). No obvious signal is observed, suggesting that the observed signals come from the $Fe_2O_3$ layer.

To separate $V_{x,\text{thermal}}$ from the total signal with $\sin(\theta)$ dependence, the geometry factor needs to be considered. The measured SMR voltage is proportional to the length of the measured area [31]

$l$, which is the length $l$ between the Hall bar channel for the longitudinal case, and the width of the Hall bar $w$ for the transverse case. Meanwhile, the thermal contribution is also proportional to the measured area [32], and as a result, there is:

$$\frac{V_{x,\text{SMR}}}{V_{y,\text{SMR}}} = \frac{V_{x,\text{thermal}}}{V_{y,\text{thermal}}} \quad (3),$$

with the ratio of the field-like torque induced voltage ($V_{x,\text{SMR}}$ and $V_{y,\text{SMR}}$), the contribution of thermal effect at longitudinal direction can be calculated with the thermal effect at transverse direction.

Figure 3(a) shows the fitting result of the $\sin(\theta)$ dependent signal for the longitudinal (black) and transverse (blue) measurement at different fields with fitting errors. Since the thermal contributions are proportional to the magnetization induced by the magnetic field, both longitudinal and transverse signals should have the same linear trend [29,30]. The goodness of fitting is estimated to be $R^2=0.94$ ($R^2=0.99$) for longitudinal (transverse) measurements, respectively. The field-like torque term is suggested to be proportional to $1/H_{\text{ext}}$ in antiferromagnetic systems[29]. Figure 3(b) shows the linear fitting result of the field-like torque term versus $1/H_{\text{ext}}$ for the longitudinal (black, $V_{x,\text{FLT}} = \frac{V_{x,\text{SMR}} H_{\text{FL}}}{2H_{\text{ext}}}$) and transverse (blue, $V_{y,\text{FLT}} = \frac{V_{y,\text{SMR}} H_{\text{FL}}}{2H_{\text{ext}}}$) measurements under different magnetic fields, respectively. The longitudinal and transverse signals have the same trend, and the signal ratio $\frac{V_{x,\text{SMR}}}{V_{y,\text{SMR}}}$ is ~4.3, which is slightly larger than the ratio ~3.5 in the mask due to sidewall etching and reveals the real geometric ratio of the device. First harmonic SMR signals in Figure 2(c) have given out a similar geometric ratio (~4.4) of the device. The goodness of the fitting is also estimated for longitudinal ($R^2=0.89$) and transverse ($R^2=0.87$) field-like torque terms respectively.

With the geometric ratio, contribution of the UMR signal can be obtained. Figure 4(a) shows the UMR contribution after the removal of the thermal contribution. The resulting UMR have a negative value compared with the thermal effect, which is different from all the other UMR or USMR behaviors observed in ferromagnetic materials. The UMR reaches ~120 µΩ with a current of 5 mA, which is 0.026% per $10^{10}$

A/cm$^2$ current density compared to the longitudinal first harmonic resistance. The value is small compared to the UMR and USMR in ferromagnetic system, but comparable to the UMR in antiferromagnetic systems.

Recent work of UMR in metallic antiferromagnetic stacks [25] has given out a possible explanation to the existence of UMR in α-Fe$_2$O$_3$. Existence of interfacial Rashba coupling can produce a nonlinear spin current due to spin-momentum locking and the symmetric distribution of electrons in the momentum space. With an applied magnetic field, the distortion of Fermi contours is significantly increased due to the existence of the spin canting [25], and the level of the distortion is proportional with the applied field, which leads to an observable UMR value. Since both Rashba coupling and band distortion happen at the interface, the existence of UMR in insulator system can be well answered. The negative UMR value may come from the dominance of inner band, which depends on the position of Fermi level and band structure at the interface. Other contributions are also considered. Since a net moment exists in α-Fe$_2$O$_3$ and interfacial band splitting can exist at Pt/ferromagnetic interface, as well as the field dependent trend of the UMR signal, it is common to consider the magnetic proximity effect (MPE) as the major contribution. However, previous study of MPE in α-Fe$_2$O$_3$ bilayers shows that MPE is not obvious until ultra-low temperature (~5 K) [33]. There is no evidence in antiferromagnet/heavy metal bilayers that the spin canting induced moment can induce MPE in the heavy metal layer. As the UMR increases almost linearly with magnetic field, another explanation may be the bilinear magnetoelectrical resistance (BMER), where the magnetoresistance changes linearly with both magnetic and electrical fields [34-36]. However, the control measurement with 5 nm Pt on Al$_2$O$_3$ substrate does not observe any signal, as shown in Figure 2(e), suggesting that the α-Fe$_2$O$_3$ layer is required for the existence of UMR.

Recently, an independent report of USMR in α-Fe$_2$O$_3$/Pt has also drawn our attention [37]. The unidirectional magnetoresistance is observed in the same material system with our observations. The authors conclude the unidirectional magnetoresistance as a USMR behavior induced by thermal random field induced imbalance in magnon creation and annihilation. The USMR value is always positive, and the value first increase then decrease with the field, which seems quite different with our results. Although the magnon creation and annihilation may not be a second order effect generated by the thermal random field considering the large birefringence like magnon transport in α-Fe$_2$O$_3$ [28], we still consider this explanation a reasonable model. To examine the existence of thermal random field induced magnonic contributions, we measured the UMR value in α-Fe$_2$O$_3$(10 nm)/Pt(5 nm) stacks. A thinner magnetic insulator can have better magnon conduction from both top and bottom channels [38], which may induce larger magnonic effect contributions. The results are shown in Figure 4(b). Both samples are measured under field values ranging from 6000 Oe to 50000 Oe. Smaller field may not saturate the magnetization of α-Fe$_2$O$_3$, as shown in Figure 1(c) and Figure 2(a). At high field, the 10 nm and 60 nm α-Fe$_2$O$_3$ samples do not have a significant difference in UMR value, suggesting an interfacial origin of UMR at high field. At low field (<12000 Oe), the net UMR has a positive value. At low field, magnonic USMR may be the dominant term, while at high field, the magnons vanishes and Fermi contour distortion and band splitting become the dominant term for the net UMR signal. A potential competing nature of antiferromagnetic UMR and magnonic USMR is revealed. The origin of antiferromagnet UMR behaviors needs to be explored in the future research.

In conclusion, we have observed a negative unidirectional magnetoresistance in antiferromagnetic α-Fe$_2$O$_3$/Pt bilayers with second harmonic measurement. The UMR shows a clear interfacial origin and a negative value, which can be attributed to the Rashba coupling and the Fermi contour distortion under the influence of spin canting. A potential thermal random field induced magnonic USMR is also explored in the α-Fe$_2$O$_3$(10 nm)/Pt(5 nm) sample. The UMR presents a signal change from positive to negative for the α-Fe$_2$O$_3$(10 nm)/Pt(5 nm) sample, which may reveal a potential competing nature for magnonic USMR and UMR. This work observed UMR in antiferromagnetic insulator and provided evidence of the interfacial origin for UMR in antiferromagnetic materials. This works also paved the way for two-terminal reading in antiferromagnet based spintronic devices.

Note added:

We are aware of an independent report of USMR in α-$Fe_2O_3$/Pt bilayers [37]. The results are developed independently and share many things in common. However, the UMR value in our research is negative and the value increases with field, which is different from that paper. We classified our observations as UMR, and concluded our observations to the interfacial Rashba coupling and band splitting due to the nature of the negative UMR value as reference suggests. However, the results in thinner α-$Fe_2O_3$ also shows a possibility of magnon and thermal fluctuation induced USMR, which is suggested in the other paper. The commons and conflicts suggest the contribution to UMR or USMR in α-$Fe_2O_3$ still needs investigation in the future research.

This work was supported, in part, by SMART, one of the seven centers of nCORE, a Semiconductor Research Corporation program, sponsored by the National Institute of Standards and Technology (NIST) and by the UMN MRSEC program under Award No. DMR-2011401. This work utilized the College of Science and Engineering (CSE) Characterization Facility at the University of Minnesota (UMN) supported, in part, by the NSF through the UMN MRSEC program. Portions of this work were conducted in the Minnesota Nano Center, which is supported by the National Science Foundation through the National Nano Coordinated Infrastructure Network (NNCI) under Award Number ECCS-2025124. P. Z. acknowledges support from Mathworks fellowship. J.-P.W. also acknowledges support from Robert Hartmann Endowed Chair Professorship. The authors acknowledge Chun-Tao Chou for the help on XRD measurements.

The data that support the findings of this study are available from the corresponding author upon reasonable request.

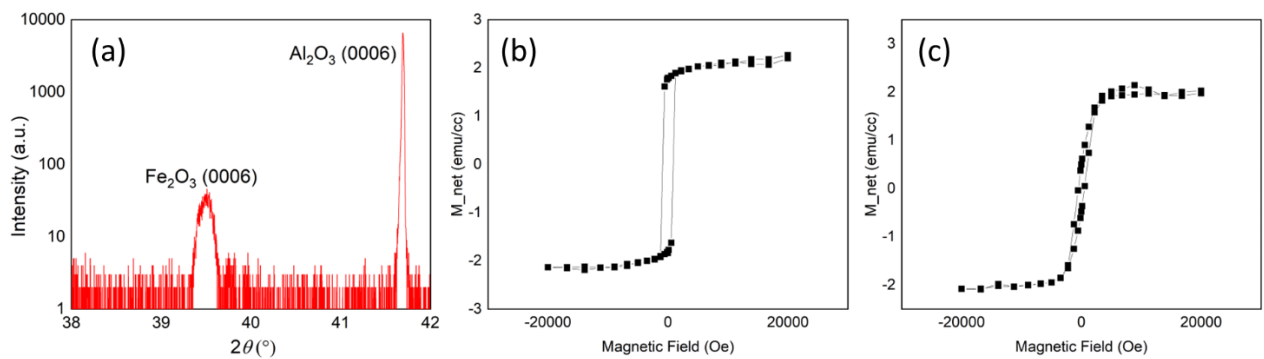

Figure 1. (a) The X-ray diffraction pattern for α-Fe$_2$O$_3$ (60 nm) sample. The SQUID measurement of easy axis [$\bar{1}2\bar{1}0$] and hard axis [$2\bar{1}\bar{1}0$] for the α-Fe$_2$O$_3$ (60 nm) sample is shown in Figure 1(b) and 1(c) respectively,

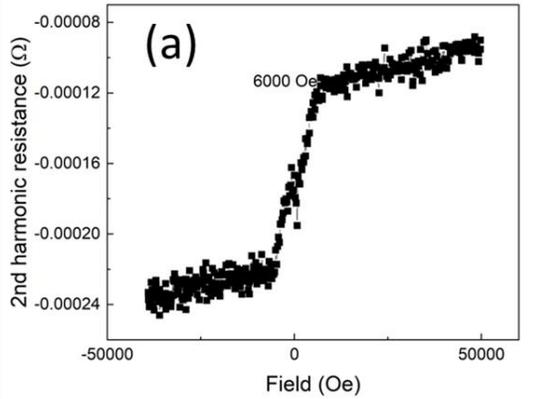
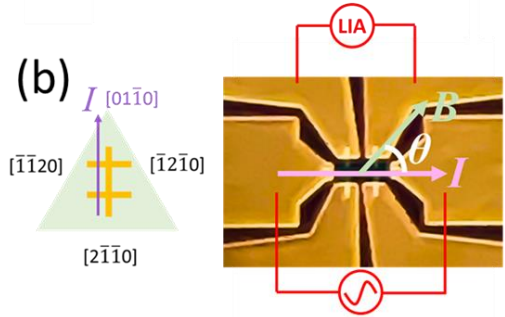
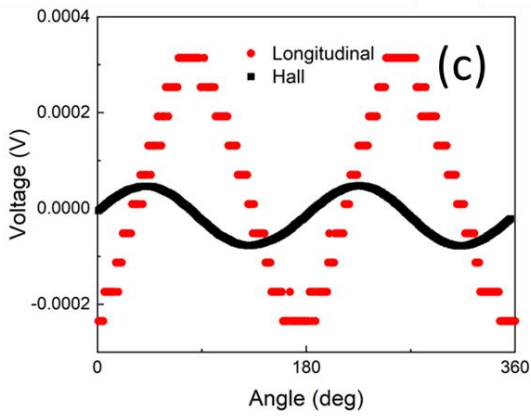
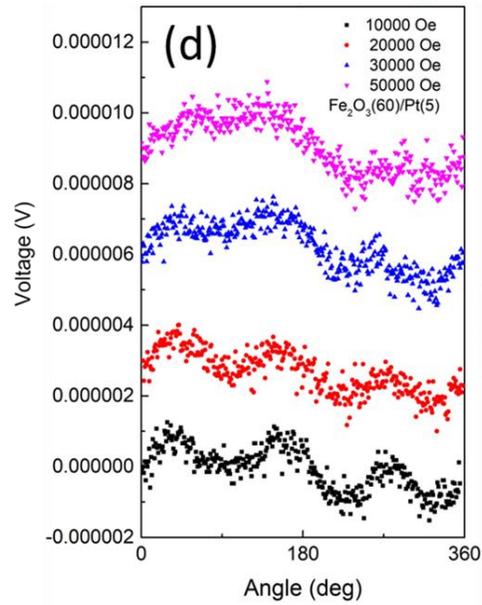
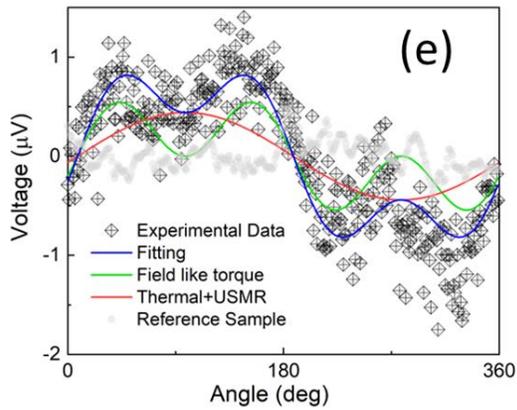
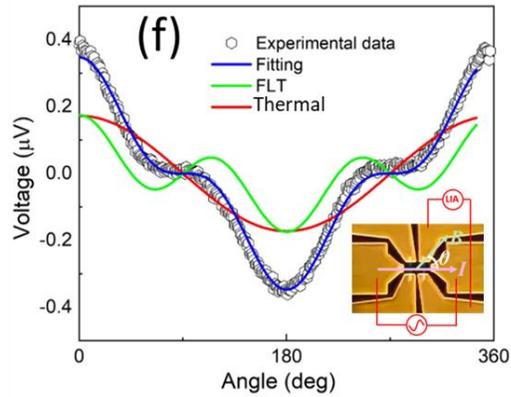

Figure 2. (a) Second harmonic Hall voltage measurement shows the hard axis direction of the sample. (b) The left section is an illustration of the sample patterning position respect to the sample crystalline direction. The right section is an illustration of the longitudinal second harmonic measurement. LIA stands for lock-in amplifier. (c) Longitudinal and Hall direction first harmonic voltage shows the SMR signal under 50000 Oe magnetic field. (d) Measured longitudinal second harmonic signals under different fields. (e) Fitting of the longitudinal second harmonic signal under 30000 Oe magnetic field. The light gray dot shows the result for the reference sample (5nm Pt on $Al_2O_3$ substrate) under the same current and field values. (f) Fitting of the transverse second harmonic signal under 30000 Oe magnetic field. The transverse measurement illustration is shown in the inner part of the figure.

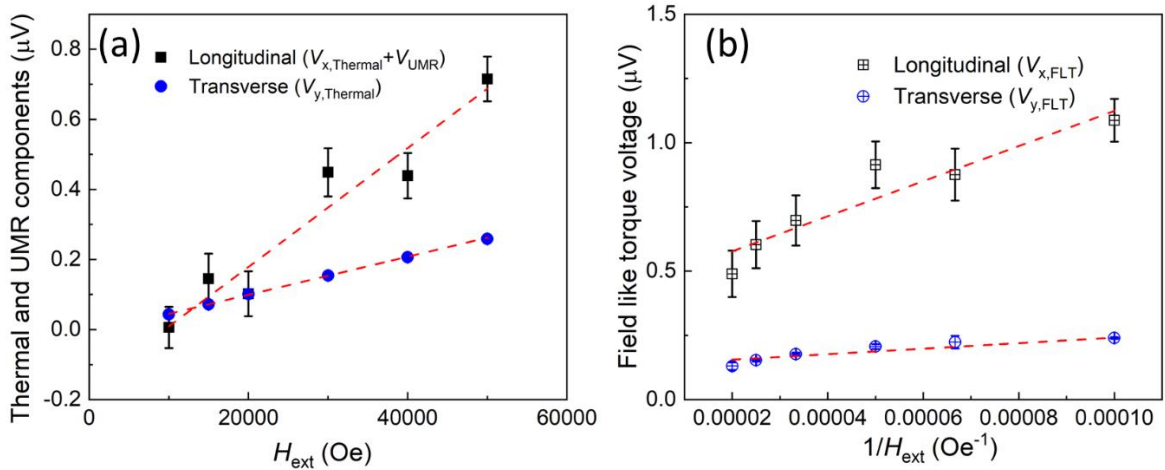

Figure 3. (a) Field dependence of the longitudinal (transverse) voltage $V_{x,\text{Thermal}} + V_{\text{UMR}}$ ($V_{y,\text{Thermal}}$). (b) Field dependence of the longitudinal (transverse) field-like torque voltage $V_{x,\text{FLT}}$ ($V_{y,\text{FLT}}$).

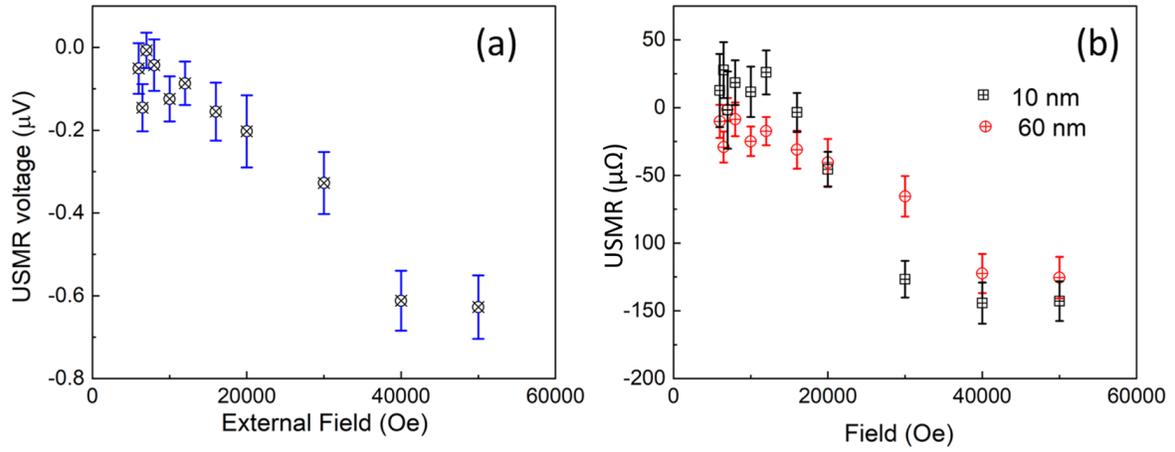

Figure 4. (a) Observed net UMR voltage in α-Fe$_2$O$_3$(60 nm)/Pt(5 nm) sample. The UMR has a negative value. (b) Observed net UMR value in α-Fe$_2$O$_3$(60 nm)/Pt(5 nm) (black square) and α-Fe$_2$O$_3$(10 nm)/Pt(5 nm) (Red circle) stacks. The UMR in α-Fe$_2$O$_3$(10 nm)/Pt(5 nm) has a signal change at ~10000 Oe, suggesting a possible competing scheme of magnonic and Rashba SOC origins.